\documentclass[12pt,amsonts,hyperref]{article}
\usepackage{graphicx}% Include figure files
\usepackage{epsfig}
\usepackage{hyperref}
\usepackage{listings}
\def\({\left(}
\def\){\right)}
\def\[{\left[}
\def\]{\right]}
\def\<{\langle}
\def\>{\rangle}

\def\b{\beta}

\def\tmp#1{}
\def\be{\begin{equation}}
\def\ee{\end{equation}}

\begin{document}

\vskip.5cm
\begin{center}
{\large \bf On the design of sell-side limit and market order tactics\footnote{The Journal of Trading, Summer 2012, Vol. 7, No. 3: pp. 29-39}}\\
\vskip.1cm
\end{center}
\vskip0.5cm

\begin{center}
{\bf
{Vladimir Markov}}
\end{center}

\begin{center}
\mbox{Liquidnet, 498 Seventh Avenue, New York, NY, 10018}
{\it vmarkov@liquidnet.com}\\
\vspace*{0.1cm}
\end{center}

\vglue 0.3truecm

\begin{abstract}
This article provides a novel framework to evaluate limit order tactics that highlights expected fill price, adverse price selection cost, and opportunity cost. We formulate the problem of optimal execution of market orders with nonlinear market impact, power law decay kernel, and stochastic and deterministic liquidity constraints. We demonstrate how these tactics can be incorporated in the uncertainty bands framework.
\end{abstract}

\section{Introduction}

In this study, our objective is to quantify general sell-side market and limit order trading tactics.

Limit orders play an important role in determining trading cost as they allow price improvement (bid-ask spread capture) and minimization of market impact. These two objectives drive the development of sell-side trading algorithms. Limit orders lower expected execution costs for the price of increased risk arising from unfavorable price movement, adverse price selection by more informed participants, failed execution and liquidation costs incurred by adhering to a trading schedule.  The liquidation cost can easily exceed the spread capture (Jeria, Schouwenaars, and  Sofianos [2009]).  There is extensive and useful research on various aspects of limit orders (Lo, MacKinlay, and Zhang [2002], Maslov [2000]).  Still, there is a lack of clarity on how to incorporate these results into actual trading tactics because the interaction between components of models is often nonlinear.  An additional hurdle is the calibration of models that often contain microscopic and  non-observable (not easily extracted from execution data) parameters.

Adverse price selection is a reality of the U.S. equity market as more than seventy percent of the volume is traded by high-frequency traders (HFTs) (Aite [2009]). The rise of HFTs and increasing market fragmentation are major trends in the U.S. equity market over the last ten years. Market fragmentation has created many loopholes that are exploited by HFTs, mostly through various forms of latency arbitrage algorithms (Nanex [2011], Arnuk S., and J. Saluzzi [2009]). Latency arbitrage algorithms impose hidden transaction costs on passive orders.

We propose a cost function to quantify the expected price (cost of non-execution) and adverse selection of limit order tactics. The backward recursion method is a convenient tool to estimate  these quantities and the results are parameterized by phenomenological macroscopic parameters that are easily estimated from trading data.

Conventionally, an optimal execution trajectory is obtained through minimization of the utility functional in the mean-variance approach. Even if the problem were exactly solvable, the optimal schedule is not uniquely defined: different definitions of risk lead to different optimal trajectories. The risk can be linear (VaR), quadratic (variance) functions of volatility, or higher statistical moments can be involved. Neglecting the opportunity cost, utility minimization leads to a trading schedule dominated by block trading at the beginning and  end of the interval. In academic (optimal execution) literature  it is assumed that liquidity is available at any given moment of time. In practice, the available liquidity fluctuates and is not consistently available.  We formulate the problem of optimal execution of market orders with stochastic and deterministic liquidity constraints.

In the last section of this paper, we propose a way to incorporate the obtained results in the uncertainty bands framework (Markov, Mazur, and Saltz [2011]). This design of schedule-based trading strategies is well suited for the integration of dynamic order-submission tactics that trade the price impact of market orders against the opportunity costs and adverse selection inherent to limit orders.

All models are presented to demonstrate the practicality of our approach and should not be construed as the models underlying Liquidnet's trading strategies.
\section{Limit Order Tactics}
We formulate and quantify  two basic limit order tactics: the Pegging Tactic (PT) and the Post and Wait Tactic (PWT).
These tactics are formulated in \mbox{\it{quote}} time (time advances when the midpoint price changes) as it is the natural time for macroscopic limit order models. The price evolution can be modeled using the binomial tree formalism. We assume a buy order to avoid trivial cluttering of notation due to the trade direction. In quote time, all subtleties of limit order book dynamics are hidden and parameterized with the constant probability of fill $q$.

Alternatively, one can use trade time and a trinomial tree approach. The trinomial tree approach offers a microscopic view, where the price evolves in trading time and the modeling of the dynamics of the limit order fill requires many detailed assumptions about limit order book dynamics.   This approach requires introduction of new parameters,  and given the noisy character of financial data and difficulties with model calibration, it may not exhibit substantial predictive power. The probability of the fill is not a constant in trading time because it increases as the queue moves. The backward recursion method allows the calculation of expected price and expected adverse selection using similar recursive equations.

The performance of a tactic is often evaluated post-trade by comparing the fill price to the mid-price $n$ ticks/quotes/seconds after the execution. A key goal in effective spread capture is to be at the top of the bid/offer queue as the adverse price selection is lower for passive orders at the top of the queue than for orders at the bottom. If your bid is at the top of the queue and it is hit, then the remaining orders in the queue support the bid price you just paid.  However, if your bid at the bottom of the queue is hit, then it is very likely that the price you just paid is in the process of becoming the new offer.
Higher event processing speed allows posting at or near the top of the bid queue, sweeping lagged liquidity from slow venues using market orders and canceling limit orders in anticipation of adverse price movement.

The limit order flow also demonstrates long memory and has a power law decay kernel (Eisler, Bouchaud, Kockelkoren [2011]) but in quote time this is a second order effect and can be neglected. Often a tactic has to have a limitation on its exposed volume in clock or trade time.

\subsection*{ Pegging Tactic}
The pegging tactic (PT) pegs the limit order at the current bid price. If the quote moves up and the order is not filled then the limit order is canceled and re-posted at the new bid level. Suppose the probability to get a limit order hit during the quote lifetime is $q$, a limit order is posted at the bid price $P_0$ at time $t=0$, the current bid-ask spread is $S$, and the tactic is bounded by $N$ quote changes, meaning that at time $t=N$, after $N$ quote changes, the order is executed aggressively, crossing the bid-ask spread to the price $P_0+(N+b)S$. We introduce market order boundary (MO) condition with $b=1$ and  midpoint (MP) boundary condition with $b=1/2$. The midpoint  boundary condition better reflects various properties of limit order dynamics for intermediate values of $N$ that are otherwise overshadowed by the certainty of market order liquidation in MO boundary conditions. The liquidation condition should not be understood literally. It is a technique to make different tactics comparable to each other through various observable expected values. Ideally, the expected price is insensitive to the boundary condition.

Provided the order survives until time $T=N$, the expected price is given by
\be
\<P_{N}\>=P_0+(N+b)S ~.
\ee
The expected price at time $T=k<N$ is
\be
\<P_{k}\>= q (P_0 + kS) + (1-q) \<P_{(k+1)}\>~.
\label{ptevol}
\ee
The recursion relation is easily solved for $\<P_0\>$ and the expected implementation shortfall $\<IS_{PT}\> = P_0 -\<P_0\>$ is
\be
\<IS_{PT}\>_N=-\frac{1}{q} (1-q+ (1-q)^{N} ((1+b)q-1)) S=g_{PT}S~.
\label{ispt}
\ee
The implementation shortfall measures the cost of non-execution of a limit order.

The factor $g_{PT}(q=1/2)=-1$ for all N for MO boundary condition and  $g_{PT}(q=2/3)= -1/2$ for all N for MP boundary condition.
The limit $N\to\infty$ is given by
\be
\<IS_{PT}\>_{\infty}=-(\frac{1}{q}-1) S~.
\label{isptinf}
\ee
Equation \ref{isptinf} is a good approximation for expected shortfall in \ref{ispt} for arbitrary $N\ge 4$.

The probability of executing at time $T=n$ is
\be
P_n(q)=q\left(1- \sum_{i=0}^{n-1}  P_n \right) = q (1-q)^{n},\,\,P_N(q)= (1-q)^{N}~.
\ee
The average waiting time is
\be
\<T\>_N =\sum_{i=0}^N i  P_i(q)=(1-q) \frac{1-(1-q)^{N}}{q}~.
\ee

Given the tolerance of $N$ quote changes before liquidation, the expected number of executions is approximately $N/\<T\>_N$.  Multiply this quantity by the average size of a limit order execution to get a rough estimate of the share capacity of the pegging tactic.

To derive the variance of the implementation shortfall, it is sufficient to know the expected value of its square:

\be
\<IS_{PT}^2\>_N=\frac{1}{q^2} \[2 - 3 q + q^2 + (1 - q)^N \times \right.
\ee
$$
\left.\(-2 + q (3 -2 N+(1+b)(-1+b+2 N)q)\)\]~.
$$

For a limit order we quantify the adverse selection as the expected loss {\em vs}. the next midpoint. The nominal  spread capture $\tilde D_k$ for a limit order executed at time $t=k$,
$$
\tilde D_k = P^{mid}_{k}-P_k = \frac{1}{2} S ~,
$$
is always positive but it does not account for the adverse price selection.
The effective spread capture $D_k$ has to be seen post-trade too, comparing the execution price to the mid-price following the execution:
\be
D_k=P^{mid}_{k+1}-P_k ~.
\label{eff_spread}
\ee

The nature of adverse selection for passive orders is that the probability to execute prior to an adverse price change is higher than the probability to execute prior to a favorable price change.
To approach this problem, we partition the fill probability $q$ into favorable and unfavorable components,
\be
q=q_{up}+q_{dn} ~,
\ee
where $q_{up}$ is the probability that the limit order is filled and the next price is up (favorable) and $q_{dn}$ is the probability that the order is filled and the next price is down (adverse).

As follows from the definition \ref{eff_spread},  $D_k = \frac{3}{2}S$ on an uptick, $D_k = - \frac{1}{2}S$ on a downtick,
and the  boundary condition is $D_{N}=-a  {S}$. $a=1/2$  for the MO boundary condition and $a=0$ for the MP boundary conditions. The backward recursion equations are
\be
\<D_k\>=-q_{dn} \frac{1}{2} S+q_{up}  \frac{3}{2} S  +(1-q_{up}-q_{dn})\<D_{k+1}\>~.
\ee
Solving this equation, we obtain the expected effective spread capture at time $t=0$ given termination time $T=N$,
\be
\<D_0\>_N=S \frac{\( 3 q_{up} -q_{dn}+  (1-q_{up}-q_{dn})^N ((1-2a) q_{dn}-(3+2 a)q_{up}) \) }{2(q_{up}+q_{dn})}  ~.
\label{adv_cost}
\ee
To reduce the probability of an adverse fill, it is critically important to be able to predict the quote change direction and cancel the limit order in advance of an adverse quote change.The probability asymmetry $q_a = q_{dn} - q_{up}$ serves as a quantitative measure of the informed part of the order flow. In terms of $q$ and $q_a$,
\be
\<D_0\>_N=\frac{1}{2 q} \( q-2 q_{a} -(1-q)^N ((1+2 a)q-2q_{a})  \) S=h_{PT} S ~.
\label{d01}
\ee
A straightforward calculation gives
\be
\<D_0^2\>_N=\frac{S^2}{4q}\(5 q - 4 q_{a}+ (1 - q)^N((4 a^2-1)q+4 q_{a})\) ~,
\label{d02}
\ee
from which one can calculate the variance of the spread capture.

For fixed $q$ and $N$, the spread capture is a linear function of $q_{a}$,
\be
\<D_0\>_N=c_1\,  q_{a}+c_0 ~.
\ee
Adverse selection is negative spread capture (price improvement).  $\<D_0\>_N$ is negative if and only if
\be
q_{a} > \frac{q}{2} \frac{(2a+1) (1-q)^N -1}{ (1-q)^N-1} ~.
\ee
In the limit $N\to\infty$,
\be
\<D_0\>_{\infty}=\frac{1}{2} \( 1-\frac{2 q_{a}}{q}\) S ~.
\ee
$\<D_0\>_{\infty}$ is negative if and only if $q_a > \frac{1}{2}q$; equivalently, $q_{dn} > 3 q_{up}$.

Another measure of adverse selection of $M$ long/short limit order pairs can be used:
\be
\Delta=\frac{1}{M \times S} \sum_{i=1}^M (P_i^{sell}- P_i^{buy}) ~.
\ee
This measurement is related to the present model by summing the adverse selection of both sides given certainty of execution ($q=1$):
\be
\Delta = (1-2 q_{a})  ~.
\ee
$\Delta > 0$ if and only if $q_d < q_u + \frac{1}{2}$. Without any protective measures, the $q_{up}/q_{dn}$ ratio can be as low as $0.1$ that shows the illusory nature of nominal spread capture $\tilde D_k$.

In order to better understand the hidden cost of the limit orders, we designed and conducted a controlled trading experiment that demonstrates adverse price selection on limit orders posted by a zero intelligence trader in stock markets that display quotes (Markov and  Saltz [2011]). We used long/short order pairs posted on the NBBO bid/ask in a Time-Weighted Average Price trading schedule. The TWAP schedule is implementation and model independent and does not introduce any intelligence in predicting price movement. This choice serves the goal of producing reproducible results. The tactical efficiency of a batch of executed order pairs is the average net profit as a fraction of the stock's average spread: $\Delta =\< P_s-P_b \> / \< S \>$, where $P_s$ is the average execution price of the sell orders, $P_b$ is the average execution price of the buy orders, and $\< S \>$ is the average spread. Every minute we send simultaneously 100 shares buy and sell passive orders to an exchange, and SORs A and B. It resulted in the Day TWAP order with 390 child orders. If an order is not filled,  we cancel and replace it after 5 minutes. Any unexecuted portion is liquidated at the close using market orders. If a market execution tactic is able to reliably buy at the bid and sell at the offer without adverse price selection, then its efficiency  approaches 1, indicating nearly perfect spread capture that happens only if adverse selection and non-execution costs are zero. This level of efficiency is an appropriate benchmark for a market maker. For INTC ticker, the experimental values (without taking into account exchange rebates $R_{exchange}$; $R_{exchange}=\$0.0029$) were $\Delta_{exchange} = -0.94$, $\Delta_{SOR_A} =-0.86$, and $\Delta_{SOR_B} = -0.84$. The result was stable for other names (LVLT and BRO) with fast convergence to the asymptotic values. So naive posting leads to $\Delta/2 \times S\approx -0.45 S$ loss versus midpoint price due to adverse selection and non-execution costs instead of naively expected $S/2$ price improvement. Also this method allows to test an arbitrary passive limit order tactics.
%Assuming that the average U.S. trading volume (as of 2011) is 10bln shares per day, the average spread is 1.3c, and the HFT earned spread is %-0.45*Spread, fee rebate is 0.23*Spread=0.3c,  the rough upper limit on the profit of HFT due to market making can be estimated as Profit$=252* %0.013*(0.45-0.23) *10^9=\$0.72$ bln/year or less than 0.01$\%$ of capitalization of the US stock market.

\subsection*{ Post and Wait Tactic}
In the Post and Wait Tactic (PWT), a trader posts an order at the lower bid level $P_L = P_0 - K*S$ of the limit order book and waits N ticks before executing the order residual aggressively.
The boundary condition at time $t=N$ for the price level $P_{M}$ is given by
\be
\<P_{N,M}\>=\theta(P_{M}+S-P_L) (P_{M}+b S)+\theta(P_L-P_{M}) P_L~,
\ee
here $\theta(x)=1$ for $x>0$ and  $\theta(x)=0$ for $x\le 0$ and price level index $M\in[-N, N]$ . The backward recursion equation is
\be
\<P_{k,M}\>=\theta(P_{M}-P_L) (\<P_{k+1,M+1}\>+\<P_{k+1,M-1}\>)/2+
\label{ep_pwt}
\ee
$$
+\delta(P_L-P_{M}) (q P_L+(1-q) \<P_{k+1,M+1}\>)~.
$$
here $k$ is the time, and $M$ is the price level index $M\in[-k,k]$ .  The matrix of expected price values  $\<P_{k,M}\>$ can be recursively calculated for any fixed bid level $P_L$ and termination time $t=N$. The Equation \ref{ep_pwt}  can be analytically solved for any fixed $K$ and $N$. For arbitrary $K$ and $N$ we rely on numerical methods for evaluation of PWT properties.

The  expected shortfall is a function of probability of fill, the bid level $K$, the termination time $N$, and bid-ask spread $S$:
\be
\<IS_{PWT}\>=P_{0}-\<P_{0,0}\>=IS_{PWT}(q,K,N,S)=g_{PWT} S~.
\ee
The adverse selection happens every time the price is equal to $P_L$ and the order is still not filled. The boundary condition is
\be
\<D_{N,M}\>=\theta(P_{M}+S-P_L) (-a  {S})~.
\ee
The backward recursion equation is
\be
\<D_{k,M}\>=\theta(P_{M}-P_L) (\<D_{k+1,M+1}\>+\<D_{k+1,M-1}\>)/2+
\label{adv_pwt}
\ee
$$
+\delta(P_{M}-P_L)(-q_{dn} \frac{1}{2} S+q_{up}  \frac{3}{2} S  +(1-q_{up}-q_{dn})\<D_{k+1,M+1}\>)~.
$$
The Equation  \ref{adv_pwt} can be analytically solved for any fixed $K$ and $N$ too and the effective spread capture is $\<D_{0,0}\>=h_{PWT} S$.

It's straightforward to implement an $\alpha$ model in this framework. An $\alpha$ model changes the probability of a next midpoint move
subject to past mid-point moves. Mean reversion models assume higher probability of a reverse tick and trend following models assume the next move in the same direction as the previous one. In order to be able to use backward evolution recurrence relationships the $\alpha$ model has to be reversed in time.The $\alpha$ is defined as an excessive return or price improvement over the naive expected price
\be
\Omega_{\alpha}=\<P_0\>-\<P_0\>_{\alpha}~.
\ee

\subsection*{Comparison of Pegging Tactic  and Post and Wait Tactic}

There are two simple ways of using a utility or cost function in the strategy design. In the uncertainty bands framework, the position relative to the middle band completely defines the behavior of the strategy. A cheaper tactic (lower cost function) is used if the strategy is ahead of the schedule and more expensive tactic is used if the strategy is behind the schedule.
The cost function has contribution from the expected shortfall  $\<IS_{}\>_{}$  and effective spread capture $\<D_{0}\>_{}$
\be
\textbf{C}=-\<IS_{}\>-\rho \<D_{0}\>~.
\ee

\begin{figure}[h]
\centering
\includegraphics[width=3in]{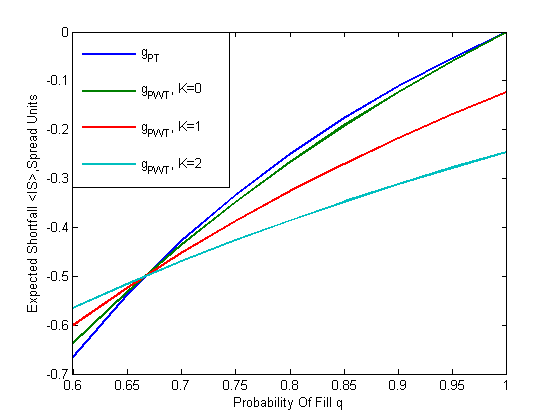}
\caption{Expected Implementation Shortfall for PT and PWT}
\label{fig:eis}
\end{figure}

\begin{figure}[h]
\centering
\includegraphics[width=3in]{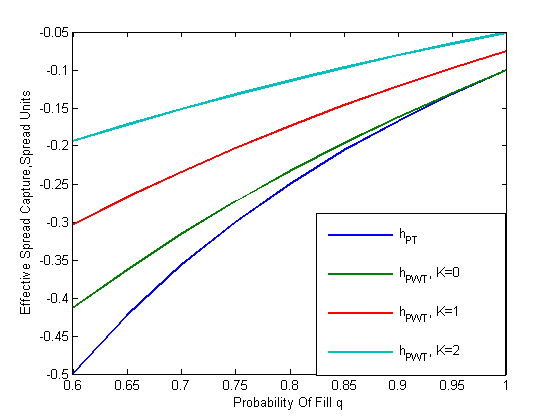}
\caption{Expected Spread Capture for PT and PWT}
\label{fig:esc}
\end{figure}

\begin{figure}[h]
\centering
\includegraphics[width=3in]{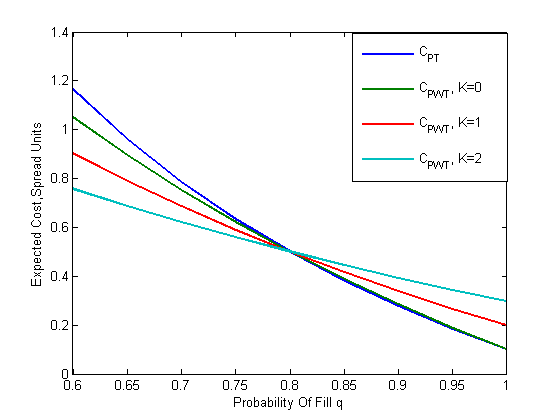}
\caption{Total Cost of PT and PWT}
\label{fig:ccc}
\end{figure}

In order to illustrate some properties of limit order tactics we compute a numerical example. For midpoint boundary condition with $b=1/2$ and  $a=0$, $q_a=0.6$ and $N=10$, the PT gives lower expected shortfall than PWT above crossing point $q_{c1}$ where $q>q_{c1}$ and lower effective spread capture for all $q$ (see Figure \ref{fig:eis} and \ref{fig:esc} ).
The total cost $\textbf{C}$ with $\rho=1$ also shows a crossing point $q_{c2}$ and is plotted on Figure \ref{fig:ccc}. The PT has the lowest total cost for $q>q_{c2}$.

The probability of fill $q$ is higher for posting at lower bid levels for PWT as the position in the queue is more favorable and does not require sophisticated technological high speed advantage. The real fill probabilities $q$ and $q_a$ are tactic and stock dependent  and can be extracted from execution data.

The MO boundary condition with $b=1$ and $a=1/2$, $q_a=0.6$ and $N=10$,  gives a uniform picture with  $g_{PT}(q)\ge g_{PWT}^{K=0}(q) > g_{PWT}^{K=1}(q) > g_{PWT}^{K=2}(q)$ , $h_{PT}(q)\ge h_{PWT}^{K=0}(q)\ge h_{PWT}^{K=1}(q)\ge h_{PWT}^{K=2}(q)$ and $C_{PWT}^{K=2}(q) > C_{PWT}^{K=1}(q) > C_{PWT}^{K=0}(q)\ge C_{PT}(q)$.

Alternatively, one can define the optimal termination time $N_T$ for a limit order by minimizing the total cost function $C_N$ with the opportunity risk included
\be
\min_N C_N=\textbf{C}+\lambda \textbf{R}~,
\ee
here risk, defined as deviation from the arrival posting price, is $R(N)=N$ for PT and $R(N)=\sqrt{N}$ for PWT. The cost function $C_N$ usually (but not always) has a well defined minimum, as normally the cost decreases and risk increases with $N$.

\subsection*{Monte-Carlo Simulation of Limit Order Tactics}
\begin{figure}[h]
\centering
\includegraphics[width=3in]{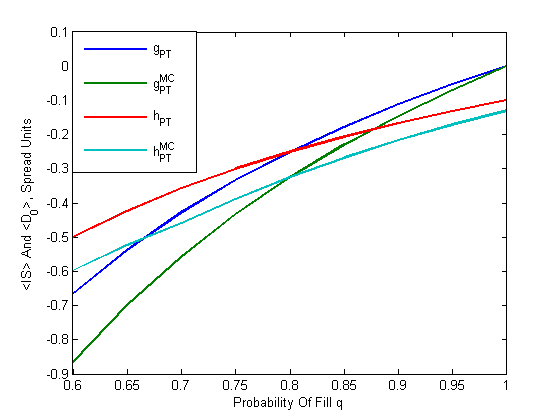}
\caption{The Monte-Carlo Simulation of PT}
\label{fig:ptmc}
\end{figure}

All parameters of the models-such as probability of fill $q$, the probability asymmetry $q_a$ and the spread $S$ - can be extracted from the execution data. It's more practical to have these parameters as distributions, not as numbers - as they change and are generally known with  limited confidence. We compute a numerical example for PT with midpoint boundary condition, $N=10$,  and discretely distributed parameters $q_{MC}\in\{(q,0.6),(q-0.1,0.2),(q+0.1,0.2)\}$, $q_a^{MC}\in\{(q_a,0.6),(q_a-0.1,0.2),(q_a+0.1,0.2)\}$, $S_{MC}\in\{(S,0.7),(2 S,0.3)\}$ . Here, the first number is the value of the parameter and the second number is the corresponding probability. We compare the results of Monte-Carlo simulation ($N_{MC}=10^4$) of PT with original PT with parameters $\{q,q_a=0.6,S,N=10\}$. The expected implementation shortfall and effective spread  capture for PT are plotted on Figure \ref{fig:ptmc}.
The uncertainty of parameters $q$ and $q_a$ and volatility of the spread decrease the expected implementation shortfall $IS^{MC}_{PT}<IS_{PT}$ and effective spread capture $D^{MC}_{PT}<D_{PT}$.

\section{Market Order Tactics}

A conventional trader's wisdom says that it is beneficial to cross the spread and trade in size if the liquidity is available at the ask.
The practical question then is  what is the minimum execution size and delay between executions given the necessity to execute $X_0$ shares during time $T$. The delay helps to avoid information leakage from the parent order and lets market volatility hide  information leakage from the execution. In econometric literature, this property is parameterized by the non-instantaneous decay kernel of the market impact. We formulate the problem of optimal execution of market orders in the light of stochastic and deterministic  liquidity constraints.
We assume a simple model that relates the mid-point price just before the trade  and the impact of the past trades.  In continuous \mbox{\it{trading}} time the price evolution is given by
\be
S_t = S_0 +\int_0^t f(\dot{x_s}) G(t-s) ds +\sigma dW_t~.
\ee
The same model in discrete \mbox{\it{trading}} time is given by
\be
S_t=S_0+\sum_{s=1}^{t-1} G(t-s) f(v_s)+\eta_s,
\label{pricedecreet}
\ee
where $ G(t-s)$  is the decay kernel, $f(v_s)=\zeta v_s^{\beta}$ is the market impact of the (buy) trade of $v_s$ shares,
and $\eta_s$ is an independent noise term that models price changes not induced by the executed trades.
We switch between two formulations freely below in order to emphasize the main points of the approach. We also assume that a  market order is implemented as a limit order with an offer price and  doesn't move the quote.
The fact that  market impact is concave means that it is beneficial to execute one large trade instead of many small ones.
The non-instantaneous decay kernel favors infrequent to frequent trades. The decay kernel is an effective way to merge two empirical facts: the long memory in market order flow and the randomness of price increments. From  a trader's point of view, the decay kernel is a simple way to parameterize
amplification of market impact due to information leakage from frequent market orders. According to (Toth, Palit,Lillo, and Farmer [2011]), the long memory in order flow originating from order splitting but not from  herding, corresponds to a sequence of  orders of a given sign originating from a single investor.

The conventional formulation of the optimal trading schedule
as minimization of the cost functional leads to the semi-trivial market order tactics: block trading at the open and  the close
of the interval and small U-shaped homogeneous trading in-between (Hasbrouck [2007]).
In case of exactly solvable exponential decay kernel, the trading schedule is block trading at the open and the close
of the interval and uniform  trading in-between (Bouchaud, Farmer, and Lillo [2008]). The block trades here are simply large blocks in \mbox{\it lit} markets, not dark block trades occurring at random times. The lack of available liquidity at the ends of an interval can easily ruin the optimality of the schedule. This means the trading schedule is superfluous if there is not enough liquidity at the offer.

%The case of linear and convex market impact is different because a non-trivial trading schedule can be defined as a
%trade-off between market memory parameterized by the decay kernel and the market impact.
The optimal solution for linear impact and a linearly decreasing decay kernel was found in (Gatheral, Schied, Slynko [2011]).

\subsection*{Optimal Trading With Liquidity Constraints}
The liquidity constraint can be implemented as a deterministic liquidity constraint, where the available offer volume is bounded by deterministic parameter $L$, or as a stochastic liquidity constraint, where the volume $v$  is a random variable distributed with probability density function $P(v)$. The latter is more attractive for practical applications. 
In this approach the order volume $X_0$ is split between two tactics: expensive uniform trading tactic (UTT) $X_u$ and cheap opportunistic liquidity taking tactic (OLTT) $X_b$
\be
X_0=X_u+X_b~.
\ee
The OLTT trades in blocks of size $v$:  the tactic with stochastic liquidity constraint  $v \ge v_0$, the tactic with deterministic liquidity constraint $v\le L$ and delay $d$ after each trade which applies to both constraints. The block trades here are simply large blocks in lit markets. The OLTT has trading cost $C_b$.
The UTT trades the residual volume $X_u=X_0-X_b$ uniformly and has trading cost $C_u$. To illustrate the approach we assume additivity of trading costs of OLTT and UTT that is justified if the OLTT has reasonably large separation between trades. The total cost of trading is given by
\be
C=C_b+C_u~.
\ee
\subsubsection*{Optimal Trading With Deterministic Liquidity Constraints}
The cost of trading an arbitrary trading schedule $u_k$ is given by
\be
C=\sum_{k=2}^{N} u_k p_k=\sum_{k=2}^{N} u_k \sum _{k'=1}^{k-1} f(u_{k'}) G(k-k')~.
\label{cost}
\ee
Using concavity of market impact, convexity of decay kernels and existence of all sums and integrals,  the general solution of optimal execution of market orders can be represented using the following ansatz
\be
u_t=\rho_b \sum_{i=1}^{N_b} \delta(t-i*d)+\rho_u , \,\,\,\,\,\  \rho_b+\rho_u \le L~,
\label{block_schedule}
\ee
here, $\rho_b$, represents the size of block trades, $N_b$ is the number of block trades, $\rho_u$ is the rate of uniform trading
and $L$ is the liquidity constraint parameter. The weak dependence of $\rho_b$ and $\rho_u$ on time is not important on a tactical level.

Assuming the $\rho_b+\rho_u \le L$ constraint and  instantaneous $G_i(k-k')=\delta(k-k')$(the sum in [\ref{cost}] has contribution only from $k'=k$), power law  $G_p(k-k')=\frac{g_p}{(k-k')^{\gamma}}$ or
exponential $G_e(k-k')=g_e e^{-(k-k') \rho}$ kernels, and the market impact function $f(v)=\zeta v^\beta$, the optimal trading schedule can be calculated. Here, $k>k'$ and $g_p$, $g_e$,$\gamma$, and $\rho$ are positive parameters of the model.

1. Instantaneous kernel $G_i(k-k')=\delta(k-k')$ with minimum delay $d_{min}$:\\
If  $\tilde X_b=L\frac{T}{d_{min}}>X_0$, $X_0$ shares should be traded in OLTT with delay $d=d_{min} {\tilde X_b}/X_0$ and size  $\rho_b=d\frac{X_0 }{T}$. If  $\tilde X_b<X_0$, the part of the order should be traded with $\rho_b$ block sized trades and the residual shares are  traded uniformly
with trading rate $\rho_u$
\be
d=d_{min} \left[\theta(X_0-\tilde X_b) 1+\theta(\tilde X_b-X_0) \frac{\tilde X_b}{X_0}\right], \,\,\, X_b=L \frac{T}{d}, N_b=\frac{T}{d},\,\,
\ee
\be
\rho_b=\min(X_0,L), \rho_u=\max(0,X_0-X_b)/T~.
\ee
We assume factorization of uniform and block trading schedules and $\rho_b \gg \rho_u$ so that we can simplify
the liquidity constraint to $\rho_b \le L$.

2. Non-instantaneous decay kernels:
\be
G_e(k-k')=g_e e^{-\rho (k-k')},\,\,G_p(t-t')=\frac{g_p}{(k-k')^{\gamma}}~.
\ee
The cost of trading every  d-th trade is given by
\be
C_b^e=\zeta \rho_b^{\beta+1}\sum_{k=2}^{N_b} \sum _{k'=1}^{k-1} g_e e^{-d (k-k') \rho} =-\zeta \rho_b^{\beta+1} g_e \frac{ N_b+e^{d \rho}(1 - e^{-d N_b \rho} - N_b)}{(e^{d \rho}-1)^2}~,
\ee
\be
C_b^p=\zeta \rho_b^{\beta+1}\sum_{k=2}^{N_b} \sum _{k'=1}^{k-1} \frac{g_e}{(dk-dk')^\gamma}=
\zeta \rho_b^{\beta+1} d^{-\gamma} \sum_{k=2}^{N_b} \sum _{k'=1}^{k-1} \frac{g_e}{(k-k')^\gamma}~.
\ee
The cost of uniform trading is
\be
C_u=\zeta \rho_u^{\beta+1}\sum_{k=2}^{N} \sum _{k'=1}^{k-1} \frac{g_u}{(k-k')^\gamma}~.
\ee
The computation of the cost of uniform trading can be accelerated using the rapidly converging continuous approximation $u_t=\rho_u$
\be
C_u\approx\zeta \rho_u^{\b+1} \int_0^N  dt \int_0^t \frac{g_p}{(t-t')^{\gamma}} dt'=\frac{\zeta \rho_u^{\b+1} g_p}{2-3\gamma+\gamma^2} N^{2-\gamma}~.
\ee
The behavior of the tactic depends on  block size $\rho_b$ and the delay between block trades $d$. The optimal value of parameters should  minimize the total cost of trading
\be
\min_{\rho_b+\rho_u \le L,d} C^{(e,p)}(\rho_b,d)=C_u+C_b^{(e,p)}, \,\,\,  X_b=\rho_b \frac{T}{d}, \,\,\, \rho_u=\max[0,\frac{X_0-X_b}{T}]~.
\ee
\begin{figure}[h]
\centering
\includegraphics[width=3in]{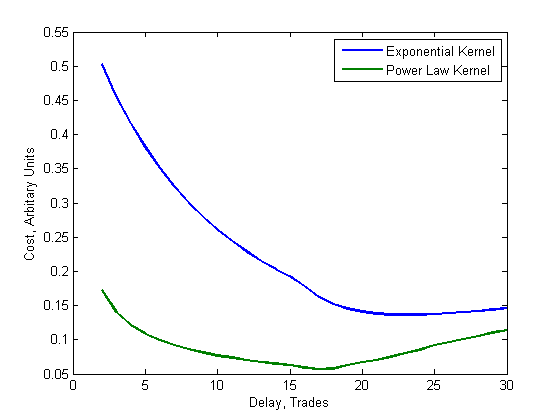}
\caption{Cost of trading for the model with deterministic liquidity constraints}
\label{fig:Cost_det}
\end{figure}
We compute a numerical example for the purpose of exploring the quantitative properties of the proposed approach.
Consider an order to buy $X_0=800$ lots of a stock during $T=500$ trades with the following model parameters:
$\beta=0.5$, $\gamma=0.5$, $\zeta=0.1$, $g_p=g_u=1$, $\rho=0.01$, $L=25$. Example of the dependence of trading cost on delay $d$ with $\rho_b=L=25$ calculated in this model are shown in Figure \ref{fig:Cost_det}. The optimal solution is given by $\{\rho_b=25, d=23\}$, $\{\rho_b=25,d=17\}$ for exponential kernel and power decay kernel, respectively. The two dimensional optimization in the parameter space $\{\rho_b,d\}$ has a tendency to hit the limit  $\rho_b+\rho_u = L$ for the execution size variable. In this case the only relevant variable is the delay $d$ between trades.

In the continuous approach, the optimal schedule can be found by  means of variational calculus as the solution of
Euler-Lagrange equation for the cost functional $C(u,\lambda,\nu)$ with constraints $\int dt   u_t=X_0$ and $u_t \le L$   implemented using Lagrange multipliers $\lambda$ and $\nu$
\be
C(u,\lambda,\nu)=\eta \int_0^T \int_0^T dt ds \theta(t-s) u_t f(u_s) G(t-s)+\lambda(\int dt   u_t-X_0)+\nu (u_t-L)~,
\ee
here, L is a liquidity constraint and  $X_0$ is the total order volume.
Using  ansatz \ref{block_schedule}, the functional minimization
can be reduced to a multidimensional function optimization.

\subsubsection*{Optimal Trading With Stochastic Liquidity Constraints}

In this framework the volume available at the offer is a random variable with probability density function $P(v)$. The optimal execution of market orders with stochastic liquidity constraints mimics closely the real execution of market orders by sell-side firms. The order volume $X_0$ is split between two tactics: uniform trading tactic (UTT) $X_u$ and opportunity taking tactic (OLTT) $X_b$.
\be
X_0=X_b+X_u~.
\ee
The OLTT trades in blocks $v \ge v_0$  with delay not less than $d$ after each trade.
The UTT trades the residual volume $X_u=X_0-X_b$ uniformly and has trading cost $C_u$. To illustrate the approach we assume additivity of trading costs for OLTT and UTT. The total cost is given by
\be
C(v_0,d)=C_b(v_0,d)+C_u~.
\ee
\subsubsection*{Monte-Carlo Simulation of Opportunity Taking Tactic}
\begin{figure}[h]
\centering
\includegraphics[width=3in]{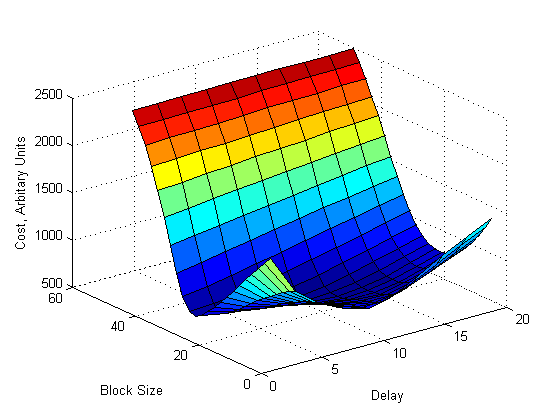}
\caption{The cost function with power law kernel}
\label{fig:costp_s}
\end{figure}

\begin{figure}[h]
\centering
\includegraphics[width=3in]{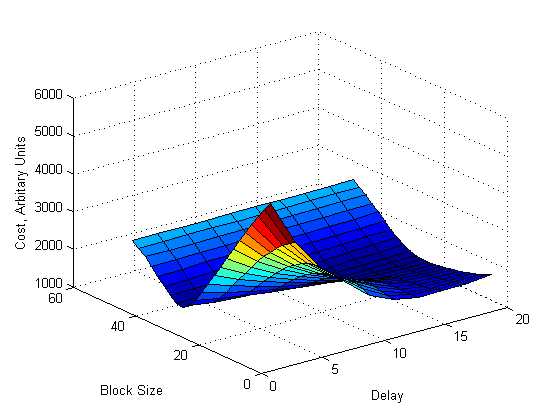}
\caption{The cost function with exponential  decay kernel}
\label{fig:coste_s}
\end{figure}

The trading cost of OLTT under different assumptions can be simulated using numerical methods.
For example, randomly drawing the volume at the offer from an unconditional $P_{u}(v)$ or
conditional $P_{c}(v_t|{v_{t'<t}})$  probability density function, a set of trade events of $T= (v_i,k_i |i\in[1 , N])$ and
a corresponding subset satisfying strategy constraints $\tilde T= T_{v \ge v_0,  k_i-\tilde k_{i-1} \ge d}$ can be generated. Here, $\tilde k_i$  is the last event with size $v\ge v_0$ and delay not less than  d.  The set of trading schedules $u_t =\sum_{k=1}^{N_b} v_k \delta (t-k)$ allow the  calculation of the expected trading cost with subset $\tilde T$
\be
C_b(v_0,d)=\<\sum_{k=2}^{N_b} u_k \sum_{k'=1}^{k-1} f(u_{k'}) G(k-k')\>_{\tilde T}~.
\ee
The decay kernel $G(t-t')$ for OLTT varies with $v_0$ threshold and $d$ delay parameters.
The kernel is instantaneous in the limit $d\gg 1$  and becomes the power law decaying kernel in the limit $d\approx 1$.
In intermediate regime, the kernel should provide an approximation between these two cases and is not well known.
We solve our earlier example to investigate this approach.
Consider an order to buy $X_0=800$ lots of shares of a stock during $T=500$ trades. The market impact function is  $f(v)=\zeta v^\beta$ with $\beta =0.5$ and $\zeta=0.1$ and the decay kernel is $G_p(k,k')=\frac{g_p}{(k-k')^{\gamma}}$ with $\gamma=0.5$ and  $g_p=1$ for power law  decay kernel
and $G_e(k,k')=g_e e^{-\rho (k-k')}$ with $\rho=0.01$ and $g_e=1$. Depending on the  selection of the unique quote procedure,  the lognormal, gamma or Weibull distribution can be used. Assuming  Weibull distribution with parameters  $\lambda_w=11.79$ and $k_w=1.21$, the cost function $C(v_0,d)$ is calculated for different delays $d$ and size thresholds  $v_0$, and $N_{MC}=10^4$ Monte Carlo samples. The two dimensional graph of the cost calculated with power law decay kernel is shown in Figure \ref{fig:costp_s} and with exponential decay kernel is shown in Figure \ref{fig:coste_s}.

The optimal solution for power law decay function is $v_0=25$  and  $d=10$ and for exponential decay function is $v_0=27$ and  $d=16$. The optimal solution for instantaneous kernel and fixed $d=16$ is $v_0=19$. All solutions give approximately the same result from the practioner's point of view.\\

\section{Combining Market and Limit Orders}

In the uncertainty bands framework, the partition of the schedule into aggressive, passive, opportunistic, and dark shares is dictated by the filled shares
position relative to the bands and the allocation among their respective execution tactics is decoupled from the schedule generation (Markov, Mazur, and Saltz [2011]).
The band separation gives the strategy discretion to wait and exploit profitable price and liquidity patterns.
When above the middle band, one can use opportunistic tactics like posting inside the limit order book with PWT and dark pool trading  with high minimum execution size as those minimize the adverse price selection and maximize the price improvement opportunities. Below the middle band, one can use a less opportunistic version of the tactics such as PT and OLTT . If the schedule falls below the lower band, one should use UTT which is the most expensive tactic to trade.

The price model \ref{pricedecreet} can be rewritten in terms of returns $r_t=p_{t+1}-p_t$ as follows
\be
r_t=G(1) \zeta_t +\sum_{s<t} k(t-s) \zeta_{s}+\eta_s,\ \zeta_t= f(v_t)~.
\ee
The decay kernel $G(l)$ decreases with $l$ and  implies that the kernel $k(l)\equiv G(l+1)-G(l)$ is negative. This means that a past series of buy trades ($\zeta_{s}>0$) tends to reduce the market impact of the subsequent buy trades and increases the market impact of the sell trades (Eisler, Bouchaud, and  Kockelkoren[2011]). The ploy here is to wait for the series of trades of the same sign  and
then execute a market order in expectation that the opposing limit orders pile up and provide a safety cushion to absorb the market impact. Alternatively, one can detect the presence of excessive liquidity at the offer and execute a market order. Both approaches are very similar and define a market order opportunity to trade a block with minimal market impact. The opportunity for execution of a limit order appears when the queue is small relative to a short-term average. Special attention must to be paid to the prevention of an execution with the adverse move of the next quote.

This leads to behavior we've dubbed the sine-cosine model wherein liquidity at the ask (sine curve) and bid (cosine) fluctuate in counter-phase.

The rebates and fees vary significantly for different exchanges and depend on client's trading volume. The rebate is a significant part of the price improvement and its management is in the domain of Smart Order Router (SOR) design.

\section{Conclusion}

We have presented a framework for the evaluation of basic market and limit order tactics. We formulate and quantify two frequently used limit order tactics: the Pegging Tactic (PT) and the Post and Wait Tactic (PWT). These tactics are formulated in quote time as it is the natural time for macroscopic limit order models. The framework highlights the expected fill price, the cost of adverse price selection and the opportunity cost.
We formulated the problem of optimal execution of market orders with nonlinear market impact, power law decay kernel and stochastic and deterministic liquidity constraints. The optimal execution of market orders with stochastic liquidity constraints mimics closely the real execution of market orders by sell-side firms. We demonstrated how these tactics can be incorporated in the uncertainty bands framework.

\section*{Acknowledgements}
The author thanks David Saltz, Vacslav Glukhov, Tito Ingargiola, and Nicola Chenosky for helpful discussions and suggestions.

\section*{References}

Aite Group Report, "New World Order: The High Frequency Trading Community and Its Impact on Market Structure", 2009.

Arnuk S., and J. Saluzzi, "Latency Arbitrage: The Real Power Behind Predatory High Frequency Trading", Themis Trading LLC White Paper, 2009.

Bouchaud, J.-P., J. D. Farmer, and F. Lillo,
``How markets slowly digest changes in supply and demand'', http://arxiv.org/abs/0809.0822.

Eisler Z., Bouchaud J.-P., and J. Kockelkoren
''Models for the impact of all order book events'', http://arxiv.org/abs/1107.3364.

Gatheral J., Schied A., and  A. Slynko,  ''Transient Linear Price Impact and Fredholm Integral Equations'', Mathematical Finance, Forthcoming,2011.

Hasbrouck. J. ''Empirical market microstructure: The institutions, economics and
econometrics of securities trading'', Oxford University Press, Oxford, 2007.

Jeria D., T. Schouwenaars, and G. Sofianos,``The all-in cost of passive limit orders'', Street Smart, Issue 38, Goldman Sachs, 2009.

Lo A., MacKinlay A., and J. Zhang, ''Econometric Models of Limit-Order Executions'',Journal of Financial Economics, vol.65,2002.

Markov V., Mazur S., and D. Saltz,''Design and Implementation of Schedule-Based Trading Strategies Based on Uncertainty Bands'', The Journal of Trading, Vol. 6, No. 4,Fall 2011.

Markov V. and D. Saltz,''Hidden cost of limit orders in the U.S. equity market'', Liquidnet Technical Report, 2011.

Maslov, S., ''Simple model of a limit order-driven market'', Physica A 278, 571, 2000.

Nanex LLC, "HFT is Out-of-Control",  Nanex White Paper, 2011.

Toth B., Palit I., Lillo F., and J. Farmer "Why is order flow so persistent?", http://arxiv.org/abs/1108.1632.

\end{document}